\documentstyle[a4wide,epsf,11pt,titlepage]{article}

\newcounter{nref}
\newcommand{\bbib}{%
  \renewcommand{\refname}{\large\bf References}%
  \begin{thebibliography}{99}%
  \setcounter{enumiv}{\thenref}}
\newcommand{\ebib}{%
  \end{thebibliography}%
  \setcounter{nref}{\arabic{enumiv}}}
\newcommand{\head}[3]{%
  \setcounter{nref}{0}%
  \thispagestyle{empty}%
  \section*{\LARGE\bf #1}%
  \stepcounter{section}%
  \addcontentsline{toc}{section}{#1}%
  \large\itshape%
  #2\\\vspace{0.1pt}\\%
  #3%
  \normalsize\upshape%
  \bigskip}

\begin{document}


\head{Prediction of nuclear reaction rates for astrophysics}
     {Thomas Rauscher}
     {Departement f\"ur Physik und Astronomie, Universit\"at Basel,\\
Klingelbergstr.\ 82, CH-4056 Basel, Switzerland}


The investigation of explosive nuclear burning in astrophysical
environments is a challenge for both theoretical and experimental
nuclear
physicists. Highly unstable nuclei are produced in such processes which
again can be targets for subsequent reactions. The majority of reactions
can be described in the framework of the
statistical model (compound nucleus mechanism, Hauser--Feshbach
approach), provided that the level density of the
compound nucleus
is sufficiently large in the contributing energy window~\cite{rauscher.rau97}.
Among the nuclear properties needed in this treatment are masses, optical 
potentials,
level densities, resonance energies and widths of the GDR.
All these necessary ingredients have to be provided in as reliable a way
as possible, also for nuclei where no such information is available
experimentally.

A recent experiment~\cite{rauscher.som98} has underlined that the low-energy
extrapolation of the widely
used optical $\alpha$+nucleus potentials may still have to be improved.
Currently, there are only few global parametrizations for
optical $\alpha$+nucleus potentials at astrophysical energies.
Most global potentials are of the Saxon--Woods form,
parametrized at energies above about 70 MeV,
e.g.~\cite{rauscher.nol87}.
The high Coloumb barrier makes a
direct experimental approach very difficult at low energies. More
recently, there were attempts to extend those parametrizations to
energies below 70 MeV~\cite{rauscher.avr94}.
Early astrophysical statistical model 
calculations~\cite{rauscher.arn72,rauscher.woo78}
made use
of simplified equivalent square well potentials and the black nucleus
approximation. Improved
calculations~\cite{rauscher.thi87} employed a phenomenological Woods--Saxon
potential~\cite{rauscher.mann78}, based on extensive 
data~\cite{rauscher.mcf66}. However, it was not clear how well all these
potentials would work for heavy targets with $A>60$ or in the
thermonuclear energy range.

Most recent experimental
investigations~\cite{rauscher.mohr94,rauscher.atz96} found a systematic 
mass-- and
energy--dependence of the optical potentials 
and were very successful in describing experimental
scattering data, as well as bound and quasi--bound states
and $B(\rm{E2})$ values,
with folding potentials.
Based on that work, a global parametrization of the volume integrals can
be found~\cite{rauscher.rau98}. In this description, the real part of
the nuclear potential is given by a folding potential $V_f(r,E)$. The imaginary
part $W(r,E)$ is of Woods--Saxon shape with a strongly energy--dependent depth.
Nuclear structure and deformation information determines the shape of
the energy--dependence by including level density dependent 
terms~\cite{rauscher.rau98}.

It is easy to show that the final transmission coefficients are not only
sensitive to the strength of the potential but also to its geometry.
Experimental data seemed to indicate that the geometry may also be
energy--dependent~\cite{rauscher.avr94,rauscher.mohr97}. At low
energies, the diffuseness of the standard volume Woods--Saxon potential
had to be set to smaller values, while the radius parameter was
increased, in order to be able to describe experimental scattering data.
This can be
understood in terms of the semi--classical theory of elastic
scattering~\cite{rauscher.bri77} which shows that the relative
importance of contributions from
different radial parts of the potential depends on the energy. 
It was shown~\cite{rauscher.bud78}
that the predicted enhanced surface absorption at low energies can be
described by an increased surface Woods--Saxon term. Thus, the
artificial change in geometry in the description of scattering data
results from the use of a volume term only.
Consequently, the
optical potential proposed here contains an imaginary part which is
given by the sum of a volume Woods--Saxon term and a surface term:
\begin{equation}
V(r,E)=V_C(r)+V_f(r,E)+i\left(W_v(E)f(r,R,a)-W_s(E)\frac{d}{dr}f(r,R,a)
\right)
\end{equation}
with
\begin{equation}
f(r,R,a)=\left[1+e^{\frac{r-R}{a}}\right]^{-1}, \quad
W_v(E)=C-\alpha e^{-\beta E}, \quad
W_s(E)=D+\gamma e^{-\delta E}\quad.
\end{equation}
The depths of the potentials are exponentially dependent on the energy,
with the volume depth $W_v$ increasing and the surface depth $W_s$
decreasing when
going to higher energies. An increasingly dominant surface term at low
energies leads to similar effects as reducing the diffuseness of a pure
volume Woods--Saxon potential. The coefficients 
are related to the height of the Coulomb barrier and the
microscopic and deformation corrections as used
in Ref.~\cite{rauscher.rau98}. The total volume integral is still given by
the relation derived in Ref.~\cite{rauscher.rau98}. The
energy--dependence of the $^{144}$Sm($\alpha$,$\gamma$)$^{148}$Gd
excitation curve~\cite{rauscher.som98} at low energies can be reproduced by 
such a description.
Nevertheless, more experimental data is needed which should be
consistently analyzed at different energies with
optical potential parametrizations similar to the one used in
Ref.~\cite{rauscher.bud78}.

Based on the well--known code SMOKER~\cite{rauscher.thi87}, an improved code for
the prediction of astrophysical cross sections and reaction rates in the
statistical model has
been developed~\cite{rauscher.oak}. Among other changes,
it includes an improved level
density description~\cite{rauscher.rau97}, updated data sets of experimental 
level information, as well as the new $\alpha$+nucleus potential. It also
allows to treat isospin effects which are especially important in
$\alpha$ capture reactions on self--conjugate target nuclei and in
proton capture reactions above the neutron separation energy. For a more
detailed presentation of the code and possible isospin effects,
see~\cite{rauscher.oak}.

\subsection*{Acknowledgements}

TR is an APART fellow of the Austrian Academy of Sciences.


\end{document}